\documentclass[lettersize,journal]{IEEEtran}
\usepackage{amsmath,amsfonts}
\usepackage{algorithmic}
\usepackage{algorithm}
\usepackage{array}
\usepackage[caption=false,font=normalsize,labelfont=sf,textfont=sf]{subfig}
\usepackage{textcomp}
\usepackage{stfloats}
\usepackage{url}
\usepackage{verbatim}
\usepackage{graphicx}
\usepackage{cite}
\usepackage{hyperref}
\usepackage{calrsfs}
\usepackage[shortlabels]{enumitem}
\begin{document}

\title{SANN-PSZ: Spatially Adaptive Neural Network for Head-Tracked Personal Sound Zones}

\author{Yue Qiao and Edgar Choueiri
\thanks{This work was supported by a research grant from Masimo Corporation. (\textit{Corresponding author: Yue Qiao.})}%
\thanks{The authors are with the 3D Audio and Applied Acoustics (3D3A) Laboratory, Princeton University, Princeton, NJ 08544, USA. (e-mail: \href{mailto:yqiao@princeton.edu}{yqiao@princeton.edu}; \href{mailto:choueiri@princeton.edu}{choueiri@princeton.edu} )}}

\markboth{Journal of \LaTeX\ Class Files,~Vol.~XX, No.~X, XX~XXXX}%
{Qiao and Choueiri: A Neural Approach For Head-Tracked Rendering of Personal Sound Zones}


\maketitle

\begin{abstract}
A deep learning framework for dynamically rendering personal sound zones (PSZs) with head tracking is presented, utilizing a spatially adaptive neural network (SANN) that inputs listeners' head coordinates and outputs PSZ filter coefficients. The SANN model is trained using either simulated acoustic transfer functions (ATFs) with data augmentation for robustness in uncertain environments or a mix of simulated and measured ATFs for customization under known conditions. It is found that augmenting room reflections in the training data can more effectively improve the model robustness than augmenting the system imperfections, and that adding constraints such as filter compactness to the loss function does not significantly affect the model's performance. Comparisons of the best-performing model with traditional filter design methods show that, when no measured ATFs are available, the model yields equal or higher isolation in an actual room environment with fewer filter artifacts. Furthermore, the model achieves significant data compression (100x) and computational efficiency (10x) compared to the traditional methods, making it suitable for real-time rendering of PSZs that adapt to the listeners' head movements.
\end{abstract}


\section{Introduction}

\IEEEPARstart{P}{ersonal} Sound Zones (PSZs) \cite{druyvesteyn1997personal,betlehem2015personal} refer to individual regions in the same space where listeners can receive different audio programs rendered using loudspeakers with minimum interference between programs. Among the PSZs, a bright zone (BZ) refers to the region where the program is delivered to the intended listener, while a dark zone (DZ) refers to the region where the program is attenuated. For a specific listener, the audio programs are categorized into either target program, which is delivered to this listener with the best possible audio quality, or interfering program, which is the target program for a different listener but may interfere with the target program for the current listener. Recent advances have been seen in the application of PSZs in various scenarios, such as automotive cabins \cite{cheer2013design,vindrola2021use,pepe2022digital}, home entertainment \cite{jacobsen2023living}, hospitals \cite{fangel2023tuning}, and outdoor spaces \cite{heuchel2020large}.

To render PSZs, digital audio filters are first generated by solving optimization problems that, given the acoustic transfer functions (ATFs) between the loudspeakers and the control points in BZ and DZ, minimize the sound energy in DZ while preserving the sound quality in BZ, then convolved with the audio programs for the loudspeaker playback. The two most commonly adopted filter design methods are acoustic contrast control (ACC) \cite{choi2002generation,galvez2015time,moller2016sound} and pressure matching (PM) \cite{poletti2008investigation,chang2012sound,vindrola2020pressure,moles2022weighted}. ACC maximizes the difference in sound energy between BZ and DZ, while PM minimizes the difference between the actual and target sound pressure in BZ and DZ. Other methods have also been proposed following ACC and PM, such as Amplitude Matching (AM) \cite{abe2022amplitude}, which relaxes the phase constraint posed in PM, and Variable Span Trade-Off Filtering (VAST) \cite{lee2020signal,brunnstrom2022variable}, which subsumes both ACC and PM as special cases and allows flexible control of the trade-off between the audio quality in BZ and the acoustic isolation in DZ. More recently, deep learning-based methods \cite{pepe2020deep,pepe2022digital,alessandri2023deep} have been developed to incorporate more flexible constraints and different filter formulations into the filter design process.

All the filter design methods mentioned above are based on the assumption that the PSZs are fixed in space and do not move with the listeners. Although theoretically, rendering large enough PSZs can provide certain flexibility for the listeners to move within the PSZs, the isolation achieved within the defined PSZ is often limited in realistic scenarios by the number of loudspeakers available, the loudspeaker placement, and the complex room acoustics. Moreover, in certain scenarios such as home entertainment \cite{jacobsen2023living}, listeners may move freely in a much larger region during the audio playback than they do in a car cabin, in which case a static PSZ setting may no longer be sufficient. Therefore, it is beneficial to develop methods that dynamically render PSZs by updating the audio filters as a function of the listener's head position. A similar need for head-tracked audio rendering has also been recognized in other loudspeaker-based applications, such as binaural audio reproduction with loudspeaker crosstalk cancellation \cite{galvez2019dynamic,ma2019concept,bruschi2021listener,kabzinski2019adaptive} and loudspeaker equalization \cite{lindfors2022loudspeaker}.


Typical approaches for updating audio filters with head tracking involve either pre-computing the filters for a discrete set of head positions for later playback \cite{ma2019concept,lindfors2022loudspeaker} or computing the filters on the fly given a new set of head positions \cite{galvez2019dynamic,bruschi2021listener,kabzinski2019adaptive}. In the former approach, the PSZ filters are dependent on the head position of multiple listeners (each with potentially six degrees of freedom) and require numerous sets to cover all possible combinations of BZ and DZ, unless the filters can be parametrized for interpolation during playback \cite{ma2019concept}, or the number of possible BZ/DZ combinations is constrained \cite{molesimplementation}; the latter approach, on the other hand, is often computationally demanding as evaluating the closed-form solutions for computing the PSZ filters (e.g., in both PM and ACC) usually involves inverting the transfer function matrices, and therefore compromises are often made to either simplify the filters with analytical system models \cite{galvez2019dynamic}, or implement adaptive solutions that are less computationally demanding \cite{vindrola2021use,bruschi2021listener,kabzinski2019adaptive,sipeiadaptive,hu2023sound,moller2024reduced}.

Leveraging the recent advances in deep learning, we propose a framework for training a spatially adaptive neural network (SANN) for dynamically rendering PSZs with head tracking. The trained SANN model takes the head coordinates as inputs and directly outputs the corresponding PSZ filter coefficients. The filter design and generation process is split into two steps: first, we train the SANN model with a dataset that contains the head coordinates of the listeners and the corresponding ATFs for the model to learn the mapping from any combination of head coordinates to the corresponding filters; then, during playback, we inference the trained model to generate the filters based on the current head positions in real-time. The idea is inspired by previous work that uses neural networks for predicting the head-related transfer functions or binaural room transfer functions for a given head position \cite{kristoffersen2021deep,richard2022deep,qiao2023neural}. To train the SANN model, we incorporate objectives used in the traditional filter design methods into a loss function and add other new constraints that are only feasible with the deep-learning approach (e.g., \cite{pepe2022digital}). In this approach, as with most existing methods, we assume that the ATFs used for filter generation are available offline (through either acoustic simulations or measurements) and do not need to be estimated in real-time. This is in contrast to approaches that use microphones for online ATF estimation \cite{sipeiadaptive,hu2023sound}.

We show that compared to the existing approaches and methods, our approach
\begin{itemize}
    \item integrates the filter generation and interpolation into a single step, as opposed to the existing method that pre-computes the filters before interpolation \cite{ma2019concept}, which greatly facilitates the implementation;
    \item achieves significant data compression and computational efficiency, which circumvent the computational bottleneck of the traditional methods and the need to implement adaptive solutions for real-time rendering;
    \item allows flexible definitions of the filter design constraints that are difficult to achieve with traditional methods;
    \item automatically ensures the filter robustness against uncertainties through data augmentation during training, which is not straightforward in the traditional approaches where ensuring robustness requires either proper regularization \cite{elliott2012robustness,coleman2014acoustic} or statistical methods \cite{zhu2017robust,moller2019influence,zhang2023cgmm};
    \item unlike traditional methods, is customizable given specific system conditions through mixed training with both simulated and measured ATFs.
\end{itemize}

The rest of the paper is organized as follows: in Sec.~\ref{sec:static}, we review traditional filter design methods for generating static PSZs and introduce the loss function for training the SANN model; in Sec.~\ref{sec:dynamic}, we explain the architecture of the SANN model; in Sec.~\ref{sec:robustness} and Sec.~\ref{sec:customization}, we discuss the training strategies for enhancing the robustness of the SANN model against uncertainties and customizing the model with measured data, respectively; in Sec.~\ref{sec:experiment}, we describe the experimental setup and evaluation metrics; in Sec.~\ref{sec:results}, we present the results of the proposed approach and compare it with the traditional methods; in Sec.~\ref{sec:summary}, we summarize the findings and provide further implications of the results.

\section{Filter Generation for Static PSZ Setup}\label{sec:static}

In traditional methods, given a static PSZ setup, the audio filters are usually generated by solving optimization problems with defined objectives and constraints. We first review the PM and AM methods in terms of their cost functions, constraints, and closed-form solutions (if there are any), then introduce the corresponding loss function in the proposed approach. 

We consider, in a general PSZ system with a single BZ and a single DZ, $L$ loudspeakers and $M$ control points that are distributed over the specified BZ and DZ. Each loudspeaker is assigned with a complex gain (also known as driving function) $g_l(\omega)$, $l=1,2,\ldots,L$, and the sound pressure at each control point is $p_m(\omega)$, $m=1,2,\ldots,M$, where $\omega$ denotes the frequency. The ATF corresponding to the loudspeaker $l$ and the control point $m$ is denoted as $H_{ml}(\omega)$, all of which assemble an ATF matrix $\mathbf{H}(\omega) \in \mathbb{C}^{M \times L}$. The sound pressure and the loudspeaker gains are related by the following equation:
\begin{equation}
    \mathbf{p}(\omega) = \mathbf{H}(\omega) \mathbf{g}(\omega),
\end{equation}
where $\mathbf{p}(\omega) = [p_1(\omega), p_2(\omega), \ldots, p_M(\omega)]^T \in \mathbb{C}^{M \times 1}$ and $\mathbf{g}(\omega) = [g_1(\omega), g_2(\omega), \ldots, g_L(\omega)]^T \in \mathbb{C}^{L \times 1}$. For simplicity, all the variables below are implicitly defined in the frequency domain unless otherwise specified.

\subsection{Pressure Matching (PM)}
The PM method minimizes the difference between the actual and target sound pressure at the control points in BZ and DZ. A typical cost function for PM is defined as
\begin{equation}\label{eq:PM}
    \mathcal{J}_{\text{PM}} = \| \mathbf{p}_{\text{T}} - \mathbf{p} \|^2 + \beta \|\mathbf{g}\|^2 = \| \mathbf{p}_{\text{T}} - \mathbf{H} \mathbf{g} \|^2 + \lambda \|\mathbf{g}\|^2,
\end{equation}
where $\mathbf{p}_{\text{T}}$ is the target sound pressure at the control points and $\lambda = \lambda(\omega)$ is the regularization parameter. The second term is added to ensure certain robustness of the filters \cite{coleman2014acoustic} and the uniqueness of the solution when $L>M$ (the underdetermined problem). The optimal loudspeaker gains (corresponding to the frequency-domain filter coefficients of finite impulse response (FIR) filters) are given by its closed-form solution of minimizing Eq.~\ref{eq:PM}:
\begin{equation}\label{eq:PM_sol}
    \mathbf{g}_{\text{PM}}^{*} = (\mathbf{H}^H \mathbf{H} + \lambda \mathbf{I})^{-1} \mathbf{H}^H \mathbf{p}_{T},
\end{equation}
where $(\cdot)^H$ denotes the conjugate transpose, and $\mathbf{I}$ is the identity matrix.

\subsection{Amplitude Matching (AM)}
The AM method was introduced by Abe et al.~\cite{abe2022amplitude} to relax the phase minimization constraint in PM, and has been shown to achieve higher acoustic isolation in DZ and lower reconstruction error in BZ compared to PM, making it preferable for rendering PSZs with monophonic programs where the phase accuracy is less critical (e.g., speech or alert sounds in automotive cabins). The cost function for AM is given by
\begin{equation}\label{eq:AM}
    \mathcal{J}_{\text{AM}} = \| \vert \mathbf{p}_{\text{T}}\vert - \vert \mathbf{H} \mathbf{g}\vert \|^2 + \lambda \|\mathbf{g}\|^2.
\end{equation}
Due to its non-convexity, a closed-form solution for AM does not exist, and iterative methods (e.g., the one based on the alternating direction method of multipliers (ADMM) in \cite{abe2022amplitude}) are required to obtain the optimal loudspeaker gains.

\subsection{Loss Function for the Proposed Approach}
In the proposed approach, the filter design objectives are incorporated into the loss function for training the SANN model. Assuming the model outputs are a set of band-limited complex filter coefficients at discrete frequencies $\omega_1, \omega_2, \ldots, \omega_N$, the first loss term is defined as
\begin{equation}
    \mathcal{L}_1 = \frac{1}{N\cdot M_{\text{B}}} \sum_{n=1}^{N}\| \vert\mathbf{p}_{\text{T,B}}(\omega_n)\vert - \vert\mathbf{H}_{\text{B}}(\omega_n) \mathbf{g}(\omega_n)\vert \|^2,
\end{equation}
where $M_{\text{B}}$ is the number of control points in BZ, and $\mathbf{p}_{\text{T,B}}(\omega_n) \in \mathbb{C}^{M_{\text{B}} \times 1}$ and $\mathbf{H}_{\text{B}}(\omega_n) \in \mathbb{C}^{M_{\text{B}} \times L}$ are the target sound pressure and the ATF sub-matrix corresponding to the control points in BZ at frequency $\omega_n$, respectively. This is equivalent to the amplitude matching in BZ. Similarly, the second loss term for minimizing the sound energy in DZ is defined as
\begin{equation}\label{eq:L2}
    \mathcal{L}_2 = \frac{1}{N\cdot M_{\text{D}}} \sum_{n=1}^{N} \| \mathbf{H}_{\text{D}}(\omega_n) \mathbf{g}(\omega_n) \|^2,
\end{equation}
where the target pressure in DZ is assumed to be zero. The third term for limiting the filter gains is defined as
\begin{equation}
    \mathcal{L}_3 = \frac{1}{N\cdot L}\sum_{n=1}^{N} \|\max(0, \vert\mathbf{g}(\omega_n)\vert-g_{\text{max}}\cdot \mathbf{1}_L^T)\|^2,
\end{equation}
where $g_{\text{max}}$ is the maximum allowed gain amplitude, and $\mathbf{1}_L^T = [1,1,\ldots,1]^T \in \mathbb{R}^{L \times 1}$. This term is similar to the regularization term in PM and AM but allows more explicit control of the gain amplitude.

A common issue with designing filters in the frequency domain is the spread of energy in the time domain, which may lead to audible artifacts such as pre-ringing or smearing of transients. Similar to the approach in \cite{pepe2022digital}, we add a fourth loss term to enforce the compactness of the filters in the time domain. Assuming each time-domain filter $\hat{\mathit{g}}_l[n]$ has a length of $\hat{N}$ samples, the fourth loss term is defined as
\begin{equation}
    \mathcal{L}_4 = \frac{1}{\hat{N}\cdot L}\sum_{l=1}^L  \|\mathit{w} \odot (\mathit{f}*\hat{\mathit{g}}_l) \|^2,
\end{equation}
where $\mathit{w}[n]$ is a weighting function that ``shapes'' the filter impulse responses, $\odot$ denotes the element-wise multiplication, $\mathit{f}[n]$ is a bandpass filter that has cutoff frequencies at $[\omega_1,\omega_N]$, and $*$ denotes the convolution operation. The bandpass filter ensures only the impulse responses for the desired rendering frequency band are considered. The loss term essentially computes the energy of the filter impulse responses from the part that ``exceeds'' the desired shape, therefore enforcing a compact structure. In practice, bandpass filtering is performed in the frequency domain, and the results are transformed back to the time domain. The weighting function is chosen as the ``inverted'' version of a typical window function to emphasize the central part of the impulse response and suppress the tails (see Sec.~\ref{sec:experiment} for details). The overall loss function is the weighted sum of the four terms:
\begin{equation}
    \mathcal{L} = \alpha \mathcal{L}_1 + (1-\alpha) \mathcal{L}_2 + \beta \mathcal{L}_3 + \gamma \mathcal{L}_4,
\end{equation}
where $\alpha$, $\beta$, and $\gamma$ are weighting parameters used to balance the importance of the four terms.

\section{Filter Adaptation to New PSZ Positions}\label{sec:dynamic}
The loss function defined above can be used to train models for generating the filter coefficients, such as in \cite{pepe2020deep,pepe2022digital,alessandri2023deep}. However, the model's architectures in these works are not able to predict the filters for new PSZ positions. Instead, the proposed SANN model follows the multi-layer perceptron (MLP) architecture that takes the head coordinates (or the PSZ center coordinates) as inputs and outputs the filter coefficients, similar to the one used in \cite{qiao2023neural}. For simplicity, we assume two PSZs (one BZ and one DZ) located in the horizontal plane, each consisting of a circle with a radius of $a$ and a center at $(x_i, y_i)$, where $i=1$ for BZ and $i=2$ for DZ. The two center coordinates are firstly normalized to $[-1+\Delta,1-\Delta]$ by a defined range $[x_{\text{min}}, x_{\text{max}}] \times [y_{\text{min}}, y_{\text{max}}]$ as the boundaries of the rendering area, and then concatenated as model inputs. The $\Delta$ is a small positive value to mitigate the issue with $\sin$ and $\cos$ functions at the boundaries, as discussed in \cite{qiao2023neural}. Then, the inputs are passed to a positional encoding layer \cite{tancik2020fourier}, $\eta \colon \mathbb{R}^4 \rightarrow \mathbb{R}^{8(K+1)}$, which maps the normalized coordinates $\tilde{\mathbf{x}}$ to high-dimensional Fourier series with a maximum order $K$: 
\begin{equation}
    \eta(\tilde{\mathbf{x}}) = \{\sin{(2^k \pi \tilde{\mathbf{x}})}, \cos{(2^k \pi \tilde{\mathbf{x}})}\}_{k=0}^{K}.
\end{equation}
These Fourier terms are then passed to several fully connected layers with the ReLU activation function, followed by a linear layer that outputs the real and imaginary parts of the filter coefficients. 

\begin{figure*}[htbp]
    \centering
    \includegraphics[width=0.8\textwidth]{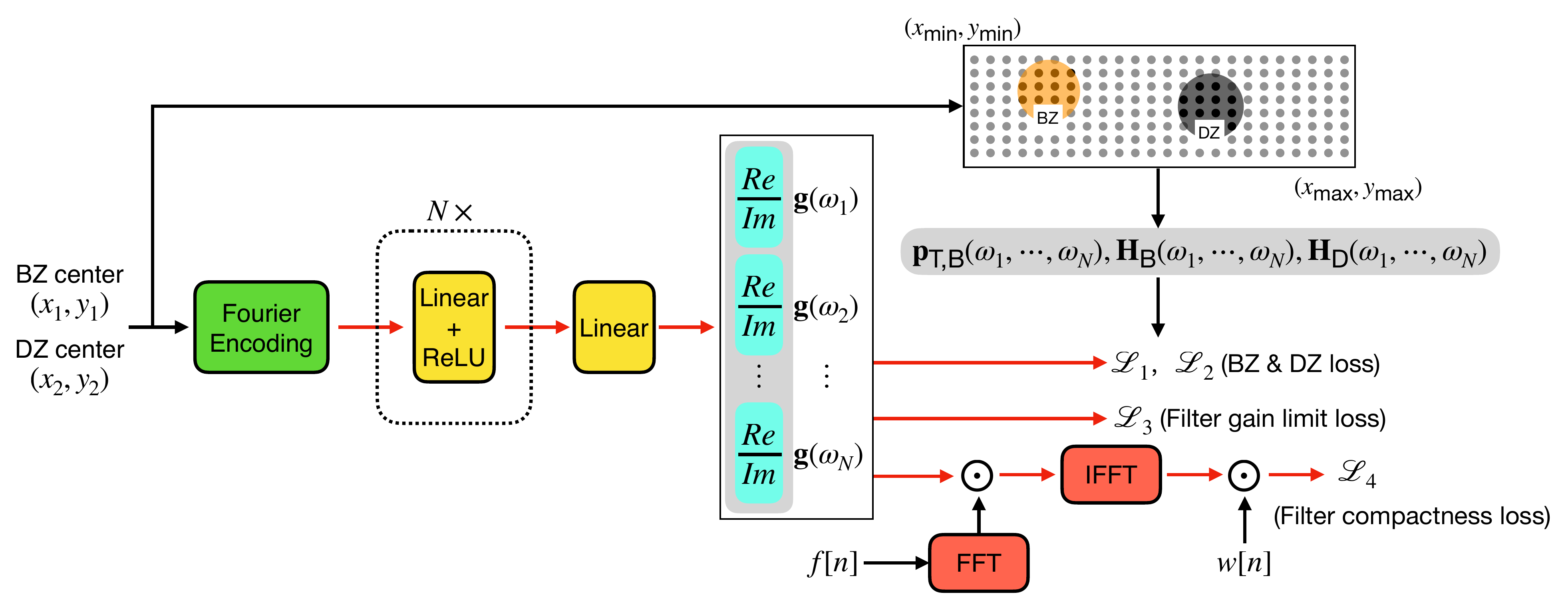}
    \caption{Illustration of the SANN architecture and the forward pass of the training process. Red arrows indicate the paths along which the gradients are back-propagated.}
    \label{fig:NNdiagram}
\end{figure*}

To train the SANN model, we assume that the ATFs in the entire rendering area are available on a uniform spatial sampling grid prior to training. During training, the center positions of the BZ and DZ are randomly sampled within the boundaries, and the spatial sampling points that fall within the defined BZ and DZ areas are selected as the control points for the loss computation. As there is no restriction on the positioning of the two zones, the loss corresponding to the DZ (i.e., Eq.~\ref{eq:L2}) is set to zero if there is overlap between the two zone areas (corresponding to the case where both listeners fall into the same BZ). Fig.~\ref{fig:NNdiagram} illustrates the model's architecture and the forward pass of the training process.

\section{Filter Robustness Against Uncertainties}\label{sec:robustness}
It is crucial to ensure the robustness of the generated filters against uncertainties in a system as the actual ATFs can deviate from the pre-obtained ones used for filter generation due to various factors. Although a certain level of filter robustness can be achieved with explicit regularization (i.e., varying the loss term $\mathcal{L}_3$), as also seen in the approaches based on traditional methods \cite{elliott2012robustness,coleman2014acoustic}, such regularization does not automatically guarantee the robustness against specific uncertainties in a PSZ system. Instead, in previous studies \cite{zhu2017robust,moller2019influence,zhang2023cgmm}, robustness constraints corresponding to different uncertainties were added to the optimization problem. Here, we propose to utilize data augmentation to implicitly guarantee robustness. Specifically, we train the model with ATFs that are randomly perturbed or modified to simulate the possible uncertainties in the system. In this way, the model learns to generate filters that are robust against the uncertainties without explicitly specifying the loss term or imposing constraints. We discuss two categories of uncertainties that are commonly seen in PSZ systems.

\subsection{System Imperfections}\label{sec:imperfections}
The first category of uncertainties comes from the imperfections or inaccuracies in a PSZ system, such as loudspeaker frequency response mismatches \cite{park2013acoustic}, loudspeaker placement errors \cite{park2013acoustic,coleman2014acoustic}, and background noise \cite{moller2019influence}. Three types of perturbations are added to the ATFs to account for these imperfections. For an ATF $H_{ml}(\omega) = A_{ml}(\omega) e^{j\phi_{ml}(\omega)}$, where $A_{ml}(\omega)$ and $\phi_{ml}(\omega)$ are the magnitude and phase of the ATF, respectively, its perturbed version is given by
\begin{equation}
    H_{ml}^{\text{pert}}(\omega) = \epsilon_{A} A_{ml}(\omega) e^{j(\phi_{ml}(\omega) + \epsilon_{\phi}(\omega))}+ \epsilon_{n},
\end{equation}
where $\epsilon_{A}$ is the multiplicative error in the magnitude that corresponds to the loudspeaker frequency response mismatch, $\epsilon_{\phi}(\omega) = k(\omega)d$ is the first-order phase error that corresponds to a loudspeaker displacement $d$ ($k(\omega)$ is the wavenumber corresponding to the frequency $\omega$), and $\epsilon_{n}$ is the additive error that corresponds to the background noise. In practice, these errors are sampled from predefined distributions on the fly during training and added to the ATF for each frequency, loudspeaker, and control point.

\subsection{Room Reflections}\label{sec:reflections}
The second category comes from the room reflections, which are inevitable in most scenarios. They can significantly affect the ATFs and eventually degrade the PSZ performance if not properly accounted for \cite{moles2022weighted}. When no in-situ acoustic measurement is available or possible, it is preferable to consider the room reflections as uncertainties to ensure the robustness of the ``universal'' filters. Instead of modeling the room reflections as errors in the ATFs, we directly generate reverberant ATFs that correspond to different room configurations for training the model.
Due to the high computational cost of simulating room reflections, such ATFs are computed pre-training with randomized room parameters (e.g., room size, reverberation time, and the positioning of the PSZ system in the room) instead of generated on the fly during training. 

\section{Filter Customization with Measured Data}\label{sec:customization}
In the previous section, the SANN model is trained with simulated ATFs to ensure its generalization capability to unknown system conditions. However, the model should also be customizable when in-situ measured ATFs are available. As actual measurements may only cover a partial region of the rendering area, we propose a training strategy that combines both simulated and measured ATFs. Specifically, we replace the simulated ATFs in the measurement region with measured ATFs and train the model with the combined dataset. The ATF perturbations as described in Sec.~\ref{sec:imperfections} are still applied to both types of ATFs. To prioritize the isolation performance in the measurement region, we assign a higher weight to the loss associated with the region. This strategy allows for both optimized performance inside the measurement region and robustness outside the region.

\section{Experimental Setup}\label{sec:experiment}
\subsection{System Configuration and Datasets}
\begin{figure}[htbp]
    \centering
    \includegraphics[width=0.3\textwidth]{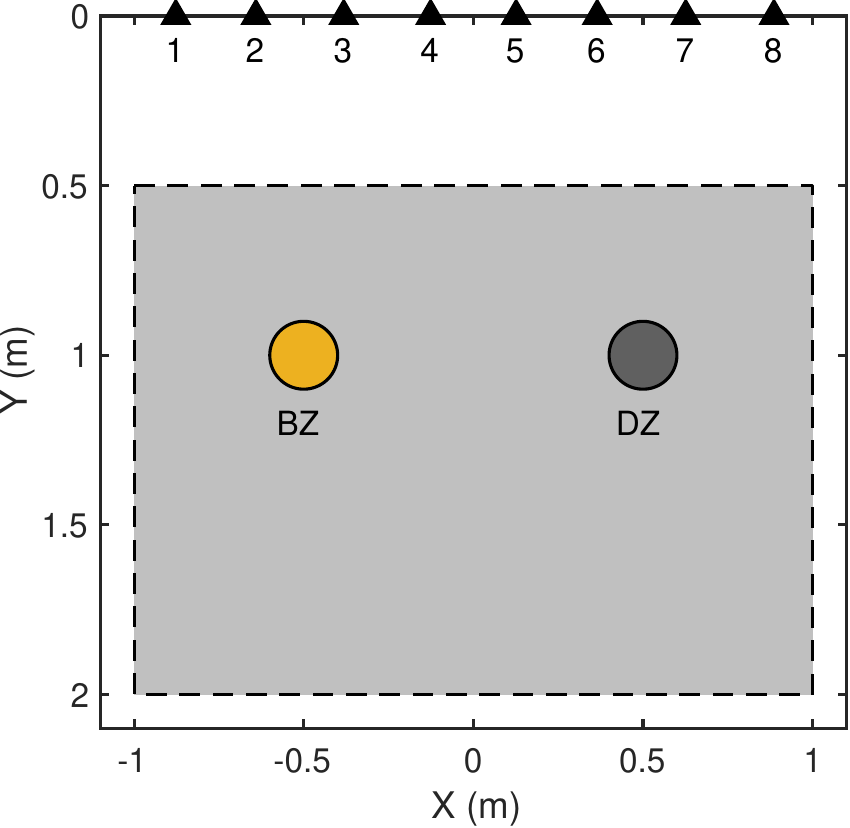}
    \caption{Illustration of the PSZ system layout. The yellow and dark gray circles represent a possible placement of BZ and DZ, respectively. The black triangles represent the transducers. The shaded rectangle indicates the rendering area within which the BZ and DZ can be positioned.}
    \label{fig:layout}
\end{figure}

The PSZ system for experimental evaluation comprises a linear array of eight loudspeakers and a rectangular rendering area with dimensions $[x_{\text{min}}, x_{\text{max}}] \times [y_{\text{min}}, y_{\text{max}}] = [-1,1] m \times [0.5,2] m$ (see Fig.~\ref{fig:layout} for illustration). The loudspeaker array layout is based on an in-house system (see \cite{qiao2024multi} for details). The PSZs are defined as horizontal circular areas with a radius of 0.1 m, approximating the upper limit of human head size.

\begin{figure}[htbp]
    \centering
    \includegraphics[width=0.4\textwidth]{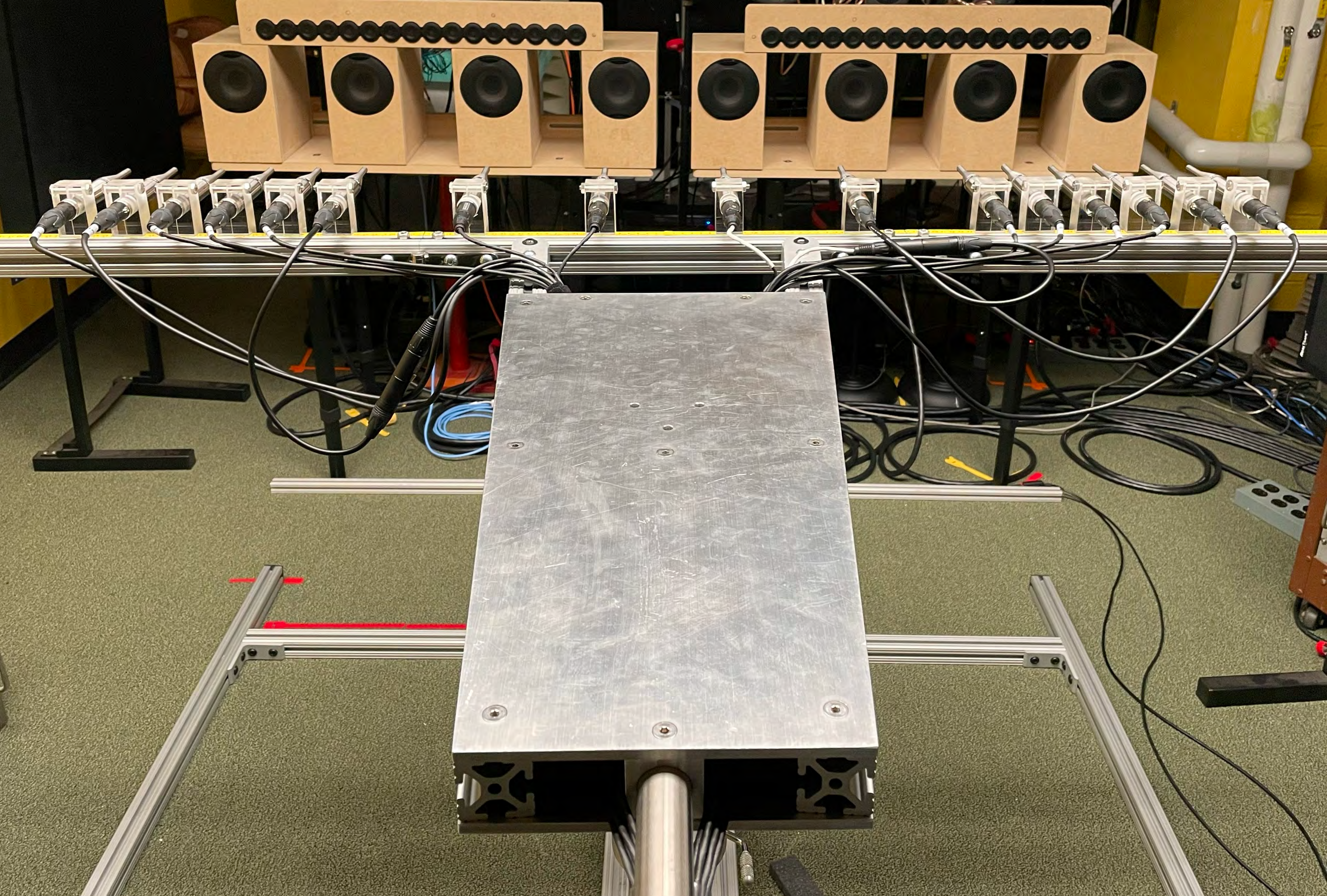}
    \caption{Photo of the measurement setup for ATFs. Note that the tweeter loudspeakers shown in the photo are not included in this work.}
    \label{fig:measurement}
\end{figure}

We utilize both simulated and measured ATF datasets in the evaluation. The simulated ATF datasets are generated within the rendering area with a spatial sampling resolution of 5 cm in both $x$ and $y$ dimensions. They are computed by treating the loudspeakers as point sources under both anechoic and reverberant conditions (the latter case is simulated with the image source method \cite{scheibler2018pyroomacoustics}). The measured ATF datasets are obtained with the in-house PSZ system in a listening room, with $\text{RT}_{60}\approx 0.24s$ calculated in the range 1300-6300 Hz. The ATFs are measured with a linear array of 16 Earthworks M30 microphones, covering the region of $[-0.84,0.84]$ m $\times [0.9,1.3]$ m, as shown in Fig.~\ref{fig:measurement}.


\subsection{Model Specifications and Training}
\begin{figure}[htbp]
    \centering
    \includegraphics[width=0.4\textwidth]{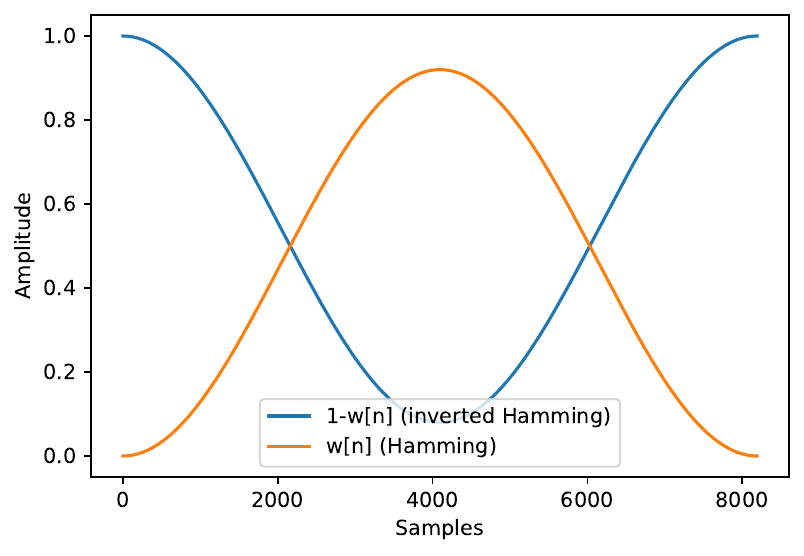}
    \caption{The weighting function $\mathit{w}[n]$ used for computing the loss term $\mathcal{L}_4$.}
    \label{fig:window}
\end{figure}

We set the model outputs to the real and imaginary parts of the frequency-domain filter coefficients, which correspond to an 8192-length impulse response at 48 kHz sampling frequency, for discrete frequencies between 100 and 1500 Hz for all eight loudspeakers. The frequency upper limit is empirically chosen as pilot experiments show that DZ is difficult to render in some areas above this frequency, potentially due to the loudspeaker spacing. The highest Fourier encoding order, $K$, is set to 3. To compute the BZ loss term $\mathcal{L}_1$, we set the target magnitude $\vert \mathbf{p}_{\text{T,B}}\vert$ as the average of the ATF magnitude responses from one edge loudspeaker and one center loudspeaker, depending on the position of the BZ (e.g., choosing the first and the fourth loudspeakers from the left if $x_1<0$). To compute the loss term $\mathcal{L}_4$, we use a 4th-order 8192-length Butterworth bandpass filter with cutoff frequencies at 100 and 1500 Hz. In the implementation, we augment the filter coefficients outside the frequency range (100-1500 Hz) with the coefficients of the bandpass filters, after normalization by the amplitude at the cutoff frequencies to avoid discontinuity in the spectrum. We take the weighting function $\mathit{w}[n]$ (shown in Fig.~\ref{fig:window}) to be an ``inverted'' Hamming window with a length of 8192. Because the filters generated in the frequency domain are non-causal by nature, we apply a circular shifting of 4096 samples to $\mathit{w}[n]$ to match the filter response in the loss computation.

All the models evaluated below have an input size of 4 (2D coordinates for BZ and DZ), an output size of 3824 (8 loudspeakers $\times$ 2 real/imaginary parts $\times$ 239 frequency bins), and three fully connected layers with 512 neurons for each layer, totaling 2.5 M parameters. A total of 10,000 random combinations of the BZ and DZ positions within the rendering area are used for training each model, with 8,000 samples for training and 2,000 samples for validation. The models are trained with the Adam optimizer with a learning rate of $10^{-3}$ and a batch size of 32 for a maximum of 400 epochs. The models are implemented with PyTorch and trained on an NVIDIA A100 GPU.

\subsection{Evaluation Metrics}
We adopt three performance metrics for evaluation, namely the inter-zone isolation (IZI), inter-program isolation (IPI), and normalized mean squared error (NMSE). Assuming two PSZs $\text{Z}_1, \text{Z}_2$ with their corresponding ATF sub-matrices as $\mathbf{H}_1$ and $\mathbf{H}_2$, we denote the filters that render BZ in $\text{Z}_1$ (or $\text{Z}_2$) and DZ in $\text{Z}_2$ (or $\text{Z}_1$) as $\mathbf{g}_1^{*}$ (or $\mathbf{g}_2^{*}$). In this work, as the filters render mono audio programs in BZ (i.e., a single vector $\mathbf{p}_\text{T}$), IZI is defined in \cite{qiao2022isolation} as
\begin{equation}
    \text{IZI}_{1} = \frac{ \|\mathbf{H}_1 \mathbf{g}^*_1\|^2 }{ \|\mathbf{H}_2 \mathbf{g}^*_1\|^2 }, \quad\quad\quad
    \text{IZI}_{2} = \frac{ \|\mathbf{H}_2 \mathbf{g}^*_2\|^2 }{ \|\mathbf{H}_1 \mathbf{g}^*_2\|^2 },
\end{equation}
where the subscript 1 (or 2) of IZI refers to the case of rendering BZ in $Z_{1}$ (or $Z_{2}$) and DZ in the other. In this particular case, IZI was shown \cite{qiao2022isolation} to be equivalent to the commonly-used Acoustic Contrast (AC) metric~\cite{coleman2014acoustic}. Correspondingly, IPI for two different BZ/DZ assignments is expressed as
\begin{equation}
    \text{IPI}_{1} = \frac{ \|\mathbf{H}_1 \mathbf{g}^*_1\|^2 }{ \|\mathbf{H}_1 \mathbf{g}^*_2\|^2 }, \quad\quad\quad
    \text{IPI}_{2} = \frac{ \|\mathbf{H}_2 \mathbf{g}^*_2\|^2 }{ \|\mathbf{H}_2 \mathbf{g}^*_1\|^2 }.
\end{equation}
Compared to IZI, which indicates the difference between BZ and DZ, IPI indicates the interference level a listener would perceive when both target and interfering programs are rendered at the same time. The NMSE between the rendered and target amplitude responses in BZ is defined as
\begin{equation}
    \text{NMSE} = \frac{1}{N\cdot M_{\text{B}}} \sum_{n=1}^{N}\frac{\| \vert\mathbf{p}_{\text{T,B}}(\omega_n)\vert - \vert\mathbf{H}_{\text{B}}(\omega_n) \mathbf{g}(\omega_n)\vert \|^2}{\|\mathbf{p}_{\text{T,B}}(\omega_n)\|^2}.
\end{equation}
All the metrics will be plotted with a logarithmic scale (i.e., taking $10\log_{10}(\cdot)$) for better visualization.

\section{Results}\label{sec:results}
We first examine the effects of loss function hyperparameters on the model's performance by evaluating it with simulated ATFs. Next, we evaluate the models with different robustness enhancement methods using measured ATFs. Furthermore, we show the results of the model customization with measured ATFs that are partially available only in certain areas. Finally, we compare the best-performing model with the traditional methods (PM and AM) for all three metrics. All results below, except for the filter response plots, are either processed by a log-weighted average function in the frequency domain, defined as
\begin{equation}
    \text{logMean}(x(\omega)) = \frac{\sum_{\omega}x(\omega)/\omega}{\sum_{\omega}1/\omega},
\end{equation} 
if shown as spatial maps or applied with a 1/6-octave smoothing \cite{tylka2017generalized} if shown as functions of frequency, after taking the logarithm.

\subsection{Effects of Loss Function Hyperparameters}\label{sec:hyperparameters}
\begin{figure}[htbp]
    \centering
    \includegraphics[width=0.35\textwidth]{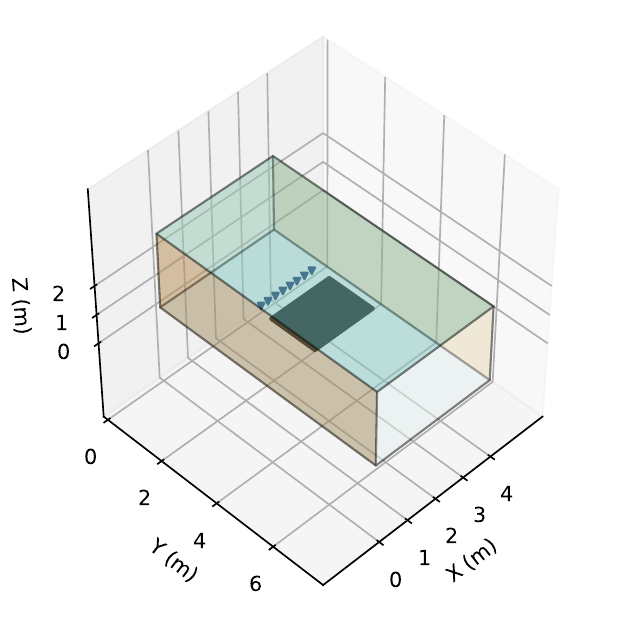}
    \caption{Illustration of the simulated shoebox room. The black triangles and rectangular area represent the loudspeakers and rendering area, respectively.}
    \label{fig:simulated_room}
\end{figure}

We first investigate the effects of the gain limit hyperparameter $g_{\text{max}}$ in the loss term $\mathcal{L}_3$ on the model's performance.
We train four models with $g_{\text{max}}$ varied and the other hyperparameters fixed. These models are trained with ATFs simulated under anechoic conditions and tested with ATFs simulated in a shoebox room (illustrated in Fig.~\ref{fig:simulated_room}) with dimensions of 4 m $\times$ 7.5 m $\times$ 2.5 m and $RT_{60}=0.24$ s; the purpose is to evaluate the performance under unknown reverberant conditions similar to those of the actual system. To evaluate the spatial dependency of the performance, we fix one zone ($\text{Z}_1$) at (-0.5, 1.0) m and vary the position of the other zone ($\text{Z}_2$) within the rendering area. For simplicity, we only show the results of $\text{IZI}_{1}$ and $\text{IPI}_{1}$ (the subscripts will be thereafter neglected) for the case where $\text{Z}_1$ is the static zone. Similarly, we only show the NMSE results for the case where $\text{Z}_1$ is BZ. 

\begin{figure*}[tbp]
    \centering
    \includegraphics[width=0.8\textwidth]{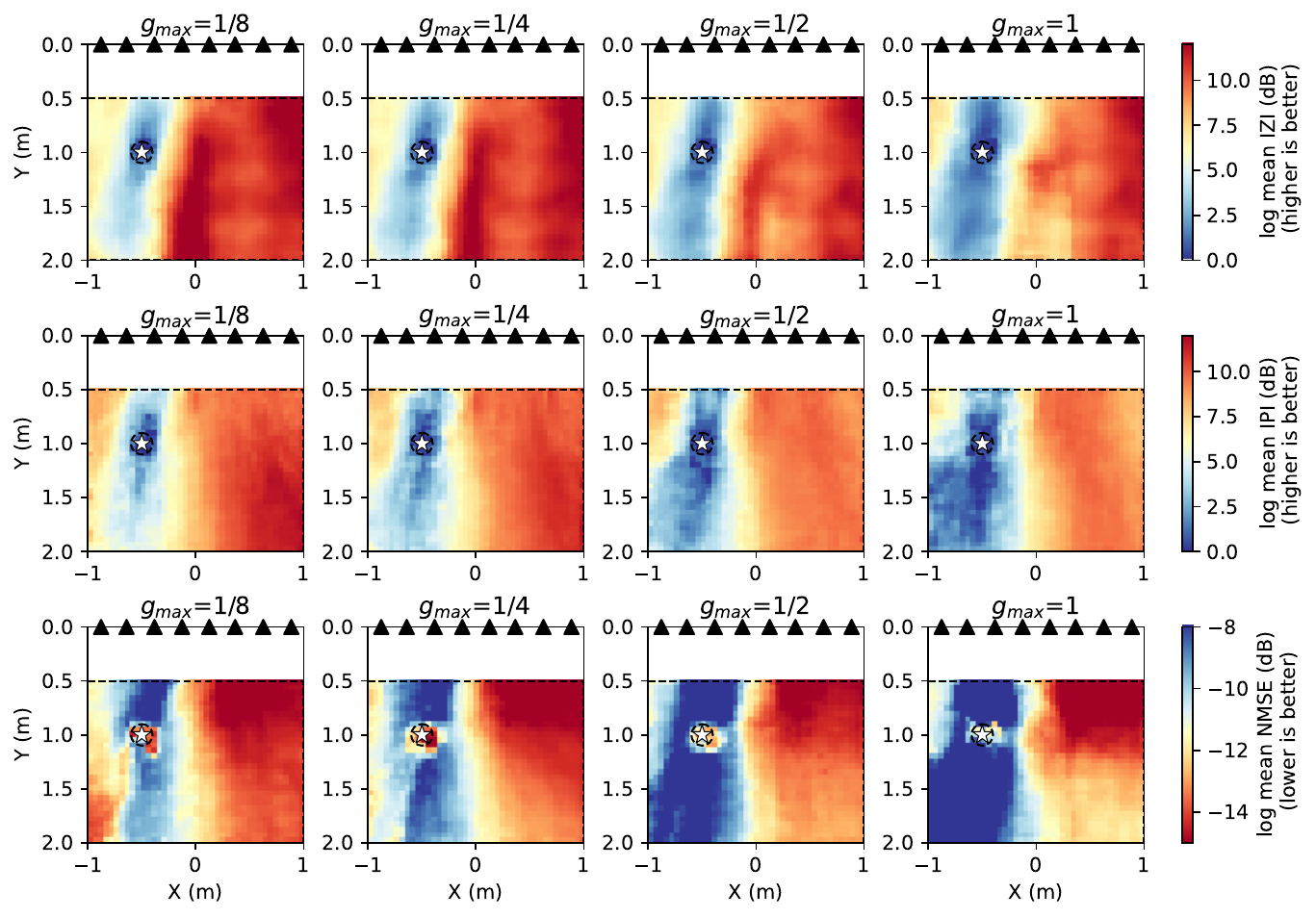}
    \caption{Spatial maps of IZI, IPI, and NMSE for models trained with different values of $g_{\text{max}}$. The columns from left to right correspond to $g_{\text{max}} = 1/8, 1/4, 1/2, 1$, respectively. The weighting parameters are set as $\alpha=0.5, \beta=0.5, \gamma=0.5$ for all models. The rows from top to bottom correspond to IZI, IPI, and NMSE, respectively. The black triangles represent the loudspeakers. The white star and the dashed circle in each map plot represent the center position and the radius of the static zone, respectively.}
    \label{fig:map_gainLimit}
\end{figure*}

Fig.~\ref{fig:map_gainLimit} shows the IZI, IPI, and NMSE results of the four models as spatial maps computed in the rendering area. First, we observe from the IZI plots on the top row that, although we set the goal of minimizing the energy in DZ regardless of its position for training, it is practically infeasible to achieve high IZI values in the regions that are close to, in front of, or behind the static BZ, due to the limited filter gains and the room reflections. Comparing the four IZI plots, we see that IZI gradually decreases as $g_{\text{max}}$ increases, especially in the regions near $x=0$ and behind the static zone. This suggests that the filters generated with a more relaxed gain limit are less robust under reverberant conditions. We also note that the region in blue (indicating low IZI values) near the static zone in all four IZI plots extends further to the left as it is further away from the static zone. This is because the size of the loudspeaker array limits the capability of rendering DZ further away from the center axis in the far field. The IPI plots in the middle row have similar trends as the IZI plots, except for minor magnitude differences due to the different definitions. The NMSE plots in the bottom row also show similar spatial patterns as the IZI plots, except when the two zones overlap. In this case, the NMSE values are relatively low because the DZ loss term is abandoned in the training. Moreover, NMSE values are higher in front of the static zone than behind it, suggesting that it is more challenging to render DZ in front of BZ than behind. This is expected as once the controlled wavefronts are formed in BZ, they are less likely to be canceled out from behind.

\begin{figure*}[tbp]
    \centering
    \subfloat{\includegraphics[width=\textwidth]{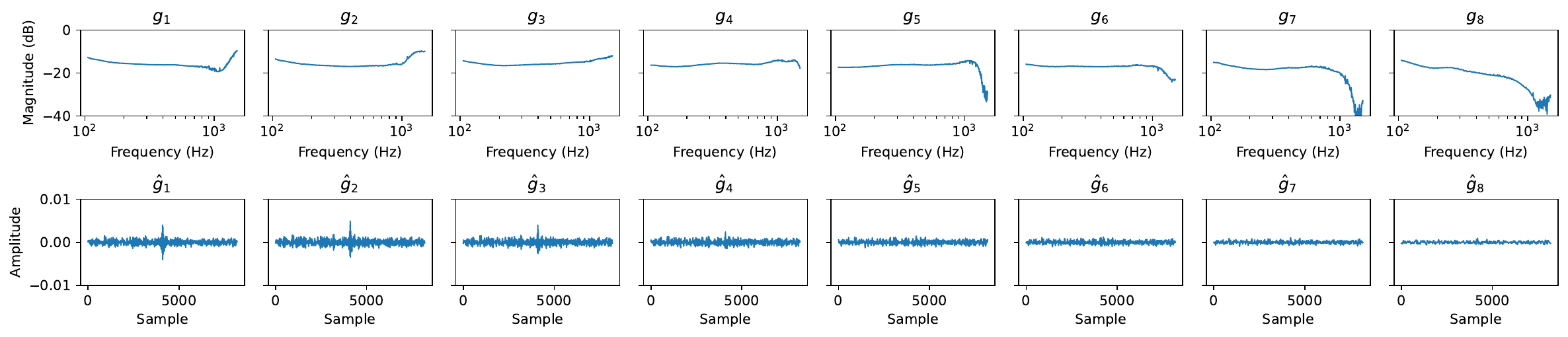}}\\ \vspace{-1em}
    \subfloat{\includegraphics[width=\textwidth]{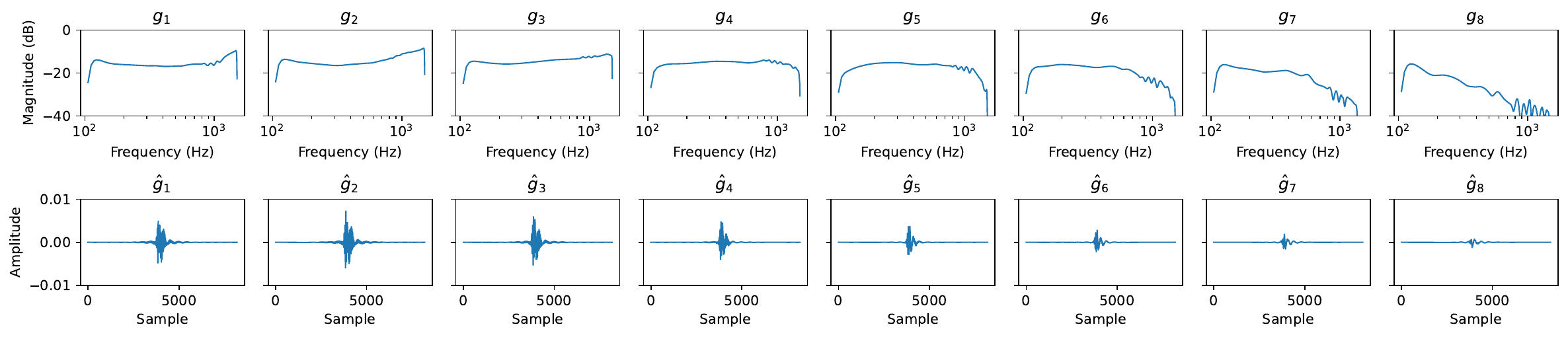}}
    \caption{Magnitude responses (between 100 and 1500 Hz) and impulse responses (after a circular shift of 4097 samples) of the filters generated by the models with $\gamma=0$ (top two rows) and $\gamma=0.5$ (bottom two rows) when the centers of BZ and DZ are set at (-0.5, 1.0) m and (0.5, 1.0) m, respectively.} 
    \label{fig:FR_IR_gamma_0}
\end{figure*}

\begin{figure}[htbp]
    \centering
    \subfloat{\includegraphics[width=0.23\textwidth]{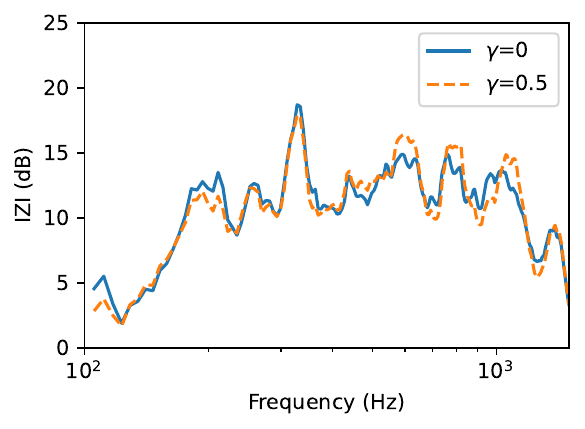}} \hfill
    \subfloat{\includegraphics[width=0.23\textwidth]{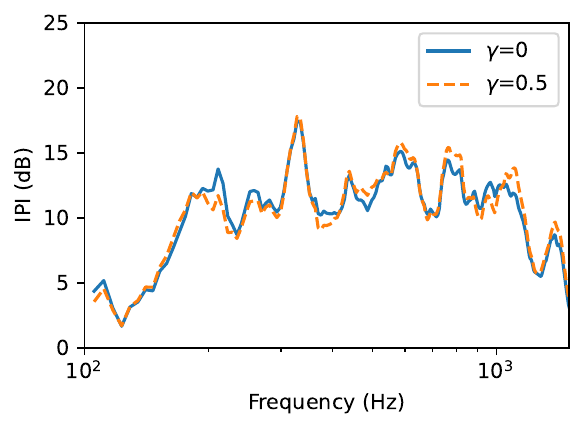}}\\ \vspace{-1em}
    \subfloat{\includegraphics[width=0.23\textwidth]{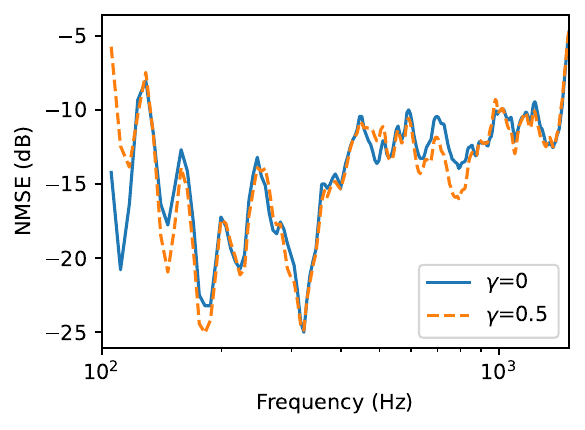}}
    \caption{Comparison of IZI, IPI, and NMSE between the models trained with $\gamma=0$ and $\gamma=0.5$, for the case where the centers of BZ and DZ are set at (-0.5, 1.0) m and (0.5, 1.0) m, respectively.}
    \label{fig:gamma_compare}
\end{figure}

Next, we investigate the effects of the weighting parameter $\gamma$ associated with $\mathcal{L}_4$, which reflects the filter compactness, on the performance. We choose the hyperparameters that correspond to the best model from the previous experiment and change $\gamma$ from 0 to 0.5. Fig.~\ref{fig:FR_IR_gamma_0} shows the magnitude responses (between 100 and 1500 Hz) and impulse responses (after a circular shift of 4097 samples) of the filters generated by the two models when the centers of BZ and DZ are set at (-0.5, 1.0) m and (0.5, 1.0) m, respectively. We see that adding the loss term $\mathcal{L}_4$ not only leads to more compact impulse responses but also better reflects the bandpass characteristics in the frequency responses. Further comparisons of the two models in terms of IZI, IPI, and NMSE (Fig.~\ref{fig:gamma_compare}) show similar levels in all three metrics, except for decreased NMSE around 100 Hz in the model with $\gamma=0.5$, which is likely due to the resulting shorter impulse responses. This suggests that adding the loss term $\mathcal{L}_4$ does not significantly affect the performance but helps to reduce the filter artifacts in the time domain.

\subsection{Performance with Simulated Training Data}\label{sec:robustnessResults}
We evaluate the real-world performance of the models trained with different strategies intended for ensuring robustness. Measured ATFs are used to determine which strategy yields the best performance given minimal knowledge of the actual system. Four models are evaluated:
\begin{enumerate}[(a)]
    \item Model trained with simulated anechoic ATFs ($\alpha=0.5,\beta=0.5, g_{\text{max}}=1/8,\gamma=0.5$); 
    \item Model trained with simulated anechoic ATFs after perturbations as described in Sec.~\ref{sec:imperfections} ($\alpha=0.5,\beta=0,\gamma=0.5$);
    \item Model trained with simulated reverberant ATFs as described in Sec.~\ref{sec:reflections} ($\alpha=0.5,\beta=0,\gamma=0.5$);
    \item Model trained with simulated reverberant ATFs after perturbations ($\alpha=0.5,\beta=0,\gamma=0.5$).
\end{enumerate}
More specifically, we model the ATF perturbations as a combination of loudspeaker displacement error $d \sim \mathcal{U}(-0.03,0.03)$ m, loudspeaker frequency response mismatch $\epsilon_{A} \sim \mathcal{U}(0.79,1.26)$ (corresponding to $\pm 2$ dB), and background noise $\epsilon_{n} \sim \mathcal{N}(0,\sigma^2)$ with $\sigma$ corresponding to 40 dB of signal-to-noise ratio. $\mathcal{U}$ and $\mathcal{N}$ denote the uniform and normal distributions, respectively. The simulated reverberant ATFs are from 51 different shoebox room configurations with randomized room geometries ($\mathcal{U}(4,7)$ m $\times \mathcal{U}(4, 7)$ m $\times \mathcal{U}(2, 4)$ m) and placement of the loudspeaker array, but the same $\text{RT}_{60}$ as the actual room. 50 room configurations are used for training and one for validation. Note that the gain limit loss term $\mathcal{L}_3$ is not included in the training of model (b)-(d) by setting $\beta=0$ for evaluating the model's performance without explicit regularization. 

\begin{figure*}[tbp]
    \centering
    \includegraphics[width=0.95\textwidth]{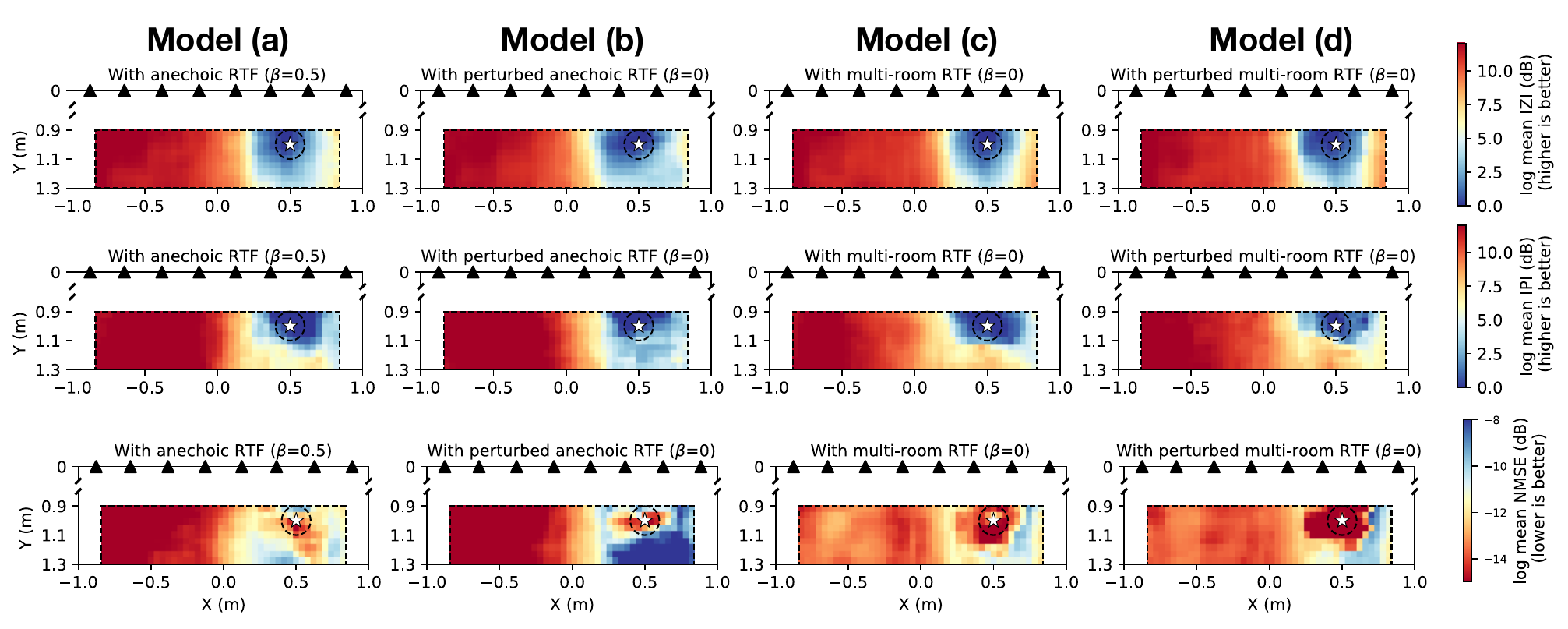}
    \caption{Spatial maps of IZI, IPI, and NMSE for the four models evaluated with measured ATFs in the region of $[-0.84,0.84] m \times [0.9,1.3] m$. The four columns correspond to models (a)-(d) as described in the text.}
    \label{fig:map_robustness}
\end{figure*}

Fig.~\ref{fig:map_robustness} shows the IZI, IPI, and NMSE results of the four models evaluated with the measured ATFs, with $\text{Z}_1$ fixed at (0.5, 1.0) m. Comparing the results of models (a) and (b), we observe that both perform similarly when the two zones are far apart (corresponding to $x\leq 0$), (b) performs clearly worse in all three metrics when $\text{Z}_2$ moves near (especially behind) $\text{Z}_1$. This indicates that (b) is less robust against the actual system uncertainties without explicit regularization and that ATF perturbations are not sufficient to ensure robustness. However, comparing (a) with (c) and (d), which are trained with reverberant ATFs without explicit regularization, we see that the models trained with reverberant ATFs lead to higher IZI and IPI when $\text{Z}_2$ is behind and to the right of $\text{Z}_1$. (c) and (d) also yield significantly lower NMSE values than (a) in the region where the two zones overlap, but at a cost of slightly higher NMSE values where $x\leq 0$. This suggests that using reverberant ATFs for training may replace the explicit regularization term and achieve higher robustness against the actual system uncertainties. Comparing (c) with (d), we see that the ATF perturbations do not significantly affect the model's performance, except for the improvement of IPI near the right boundary of the rendering area and NMSE in the overlapping region. This suggests that room reflections may be the dominant source of uncertainties in the actual system.

\subsection{Performance with Mixed Simulated/Measured Training Data}\label{sec:customization_results}
\begin{figure*}[htbp]
    \centering
    \includegraphics[width=0.95\textwidth]{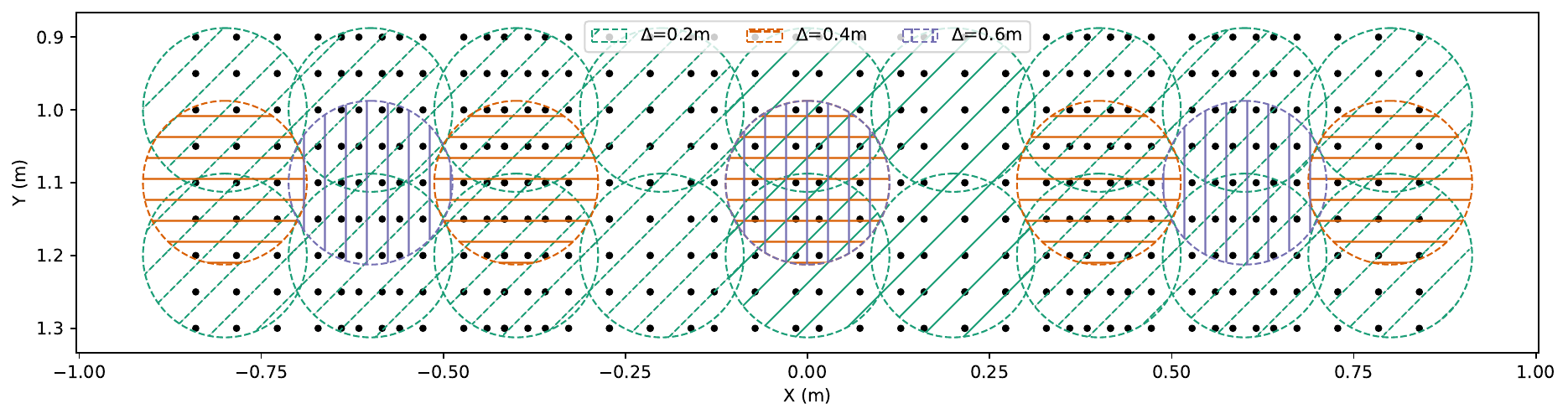}
    \caption{Illustration of the spatial sampling schemes for choosing the measured ATFs for model training. Black dots represent the spatial grid of the measured ATFs. Circles represent the areas in which the measured ATFs are selected. Different colors/hatching patterns represent different sampling schemes, for which the spacing between the centers of the circles, $\Delta$, is varied from the set of $\{0.2, 0.4, 0.6\}$ m.}
    \label{fig:sampling_scheme}
\end{figure*}

In contrast to the previous evaluation where measured data is unavailable for model training, we train three additional models with a mixture of simulated ATFs and the measured ATFs from the previous evaluation. The amount of measured ATFs in the training data is varied for each model by spatially sampling the measured ATFs with different spacings, as illustrated in Fig.~\ref{fig:sampling_scheme}. As the measured ATFs are not evenly distributed in the rendering area, we sample the data by specifying circular regions with the same size as the PSZs and with spacings of $\Delta = \{0.2, 0.4, 0.6\}$ m between the centers of the circles and selecting the measured ATFs that fall within the circles. The rest of the rendering area, once the measured ATFs are selected, is still filled with simulated ATFs from multiple room configurations. All three models use the same settings as model (d) in Sec.~\ref{sec:robustnessResults}, except that the weighting of the DZ loss term ($\mathcal{L}_2$) corresponding to the measured ATFs is increased by a factor of 2 to prioritize the isolation performance in the measurement region.

\begin{figure*}[tbp]
    \centering
    \includegraphics[width=0.95\textwidth]{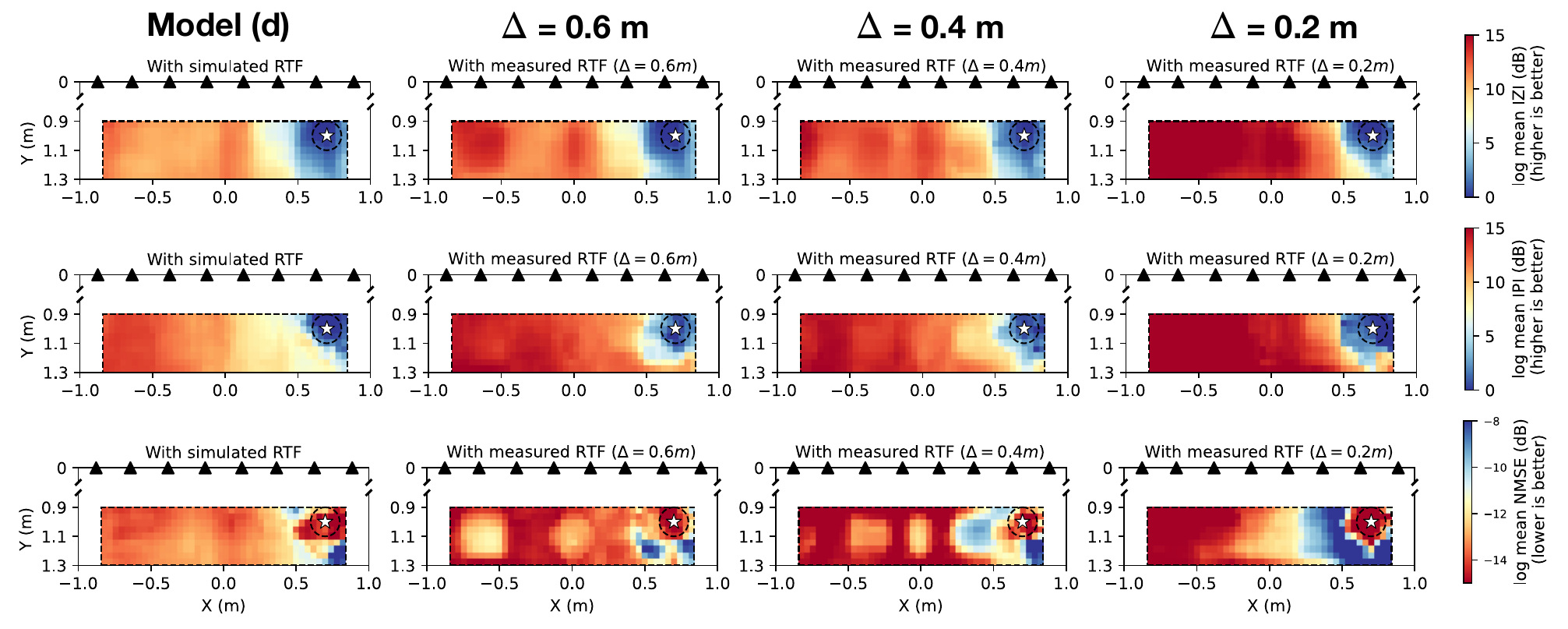}
    \caption{Spatial maps of IZI, IPI, and NMSE for the four models evaluated with measured ATFs in the region of $[-0.84,0.84] m \times [0.9,1.3] m$. The leftmost column corresponds to model (d) in Sec.~\ref{sec:robustnessResults}, and the other three columns correspond to the models trained with different amounts of measured ATFs (corresponding to $\Delta = \{0.6, 0.4, 0.2\}$ m in Fig.~\ref{fig:sampling_scheme}).}
    \label{fig:map_measuredATF}
\end{figure*}

Fig.~\ref{fig:map_measuredATF} shows the spatial maps of IZI, IPI, and NMSE for model (d) in Sec.~\ref{sec:robustnessResults} and the three models trained with different sets of measured ATFs, with $\text{Z}_1$ fixed at (0.7, 1.0) m (note the change of scale in IZI and IPI compared to Fig.~\ref{fig:map_robustness}). Comparing the results of the four models, we see a gradual improvement in IZI and IPI as more measured ATFs are included in the training data. For the cases of $\Delta=0.6$ m and $\Delta=0.4$ m, the increase in IZI mostly appears in the regions where the measured ATFs are selected because the models are designed to prioritize the minimization of DZ energy in these regions; the increase in IPI is more evenly distributed across the rendering area because the emphasized DZ loss is associated with the region near the fixed $\text{Z}_1$, therefore affecting the overall performance. This indicates that the model can still benefit from the limited measured data when the PSZ is in the vicinity of the measurement region. When measured ATFs are available for the entire rendering area ($\Delta=0.2$ m), the model achieves the highest IZI and IPI values (over 15 dB in most regions). For NMSE, we see that the models with partially available measured ATFs yield higher NMSE in the measurement regions but lower NMSE in the rest of the rendering area. This is due to prioritizing the DZ loss term (and therefore sacrificing the BZ loss term) when $\text{Z}_2$ moves into the measurement region. However, the compromised NMSE performance is still comparable to that of the baseline model (trained with simulated ATFs only). In practice, the trade-off between the isolation performance and the rendering quality can be adjusted by varying the hyperparameter $\alpha$ based on the system requirements.

\subsection{Comparison with Traditional Methods}\label{sec:method_comparison}

Lastly, we compare the best-performing SANN model with the traditional methods (PM and AM) for all three metrics. Similar to Sec.~\ref{sec:robustnessResults}, we assume that measured data is unavailable for training and use simulated ATFs for filter generation. As the robustness enhancement methods proposed in Sec.~\ref{sec:robustness} do not directly apply to the traditional methods, we use anechoic ATFs for filter generation with PM and AM and rely on explicit regularization to ensure robustness. The regularization parameter $\lambda$ in the cost functions of PM (Eq.~\ref{eq:PM}) and AM (Eq.~\ref{eq:AM}) are set to $\sigma_{\text{max}}(\mathbf{H}(\omega))\times 0.05$ for each frequency $\omega$, similar to the approach in \cite{abe2022amplitude}. $\sigma_{\text{max}}$ denotes the largest singular value of the matrix. While not shown here, we found that the filter performance with PM and AM is insensitive to the choice of $\lambda$ as long as it is within a reasonable range. The model evaluated is the same as model (d) in Sec.~\ref{sec:robustnessResults}. The AM method is implemented with the majorization–minimization algorithm in \cite{abe2022amplitude}.

\begin{figure*}[thbp]
    \centering
    \subfloat{\includegraphics[width=\textwidth]{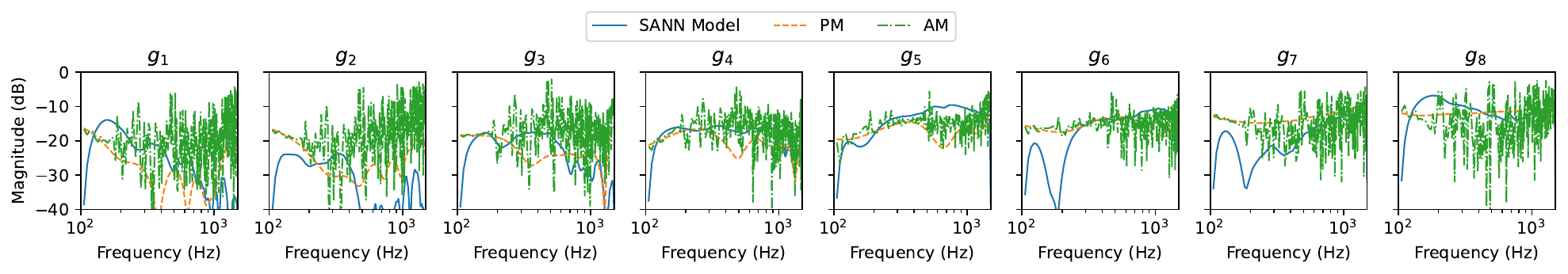}}\\ \vspace{-1em}
    \subfloat{\includegraphics[width=0.8\textwidth]{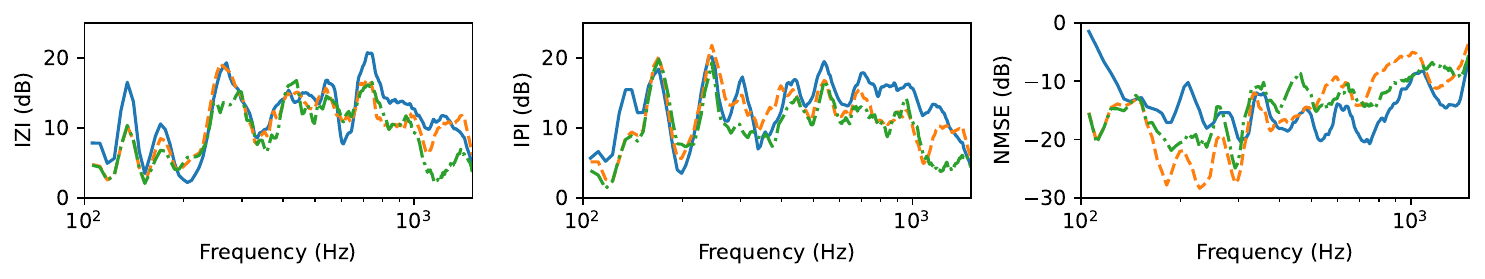}}
    \caption{Top: Magnitude responses of the filters generated by the SANN model, the PM method, and the AM method. Bottom: IZI, IPI, and NMSE results for the SANN model and the PM and AM methods.}
    \label{fig:method_comparison}
\end{figure*}

For simplicity, we only show the results for the case where $\text{Z}_1$ is fixed at (0.6, 1.0) m and $\text{Z}_2$ is at (-0.6, 1.0) m. Fig.~\ref{fig:method_comparison} shows the filter magnitude responses for the case of $\text{Z}_1$ being BZ, and the IZI, IPI, and NMSE results for the SANN model (model (d) in Sec.~\ref{sec:robustnessResults}) and the PM and AM methods. We observe from the filter magnitude response plots that the filters generated by the SANN model have a lower average magnitude at low frequencies than those generated by PM or AM; this is likely the outcome of optimizing the robustness against room reflections during model training. Moreover, the filters generated by AM have a significantly larger variance in magnitude than the other two approaches as phase is not constrained in the optimization. For the performance metrics, we see that the SANN model has equal or better (especially below 200 Hz and around 1000 Hz) IZI and IPI performance than PM and AM. It also yields lower NMSE than PM and AM above 300 Hz but higher NMSE below 300 Hz, which is likely due to the filter compactness enforced by the loss term $\mathcal{L}_4$, as discussed in Sec.~\ref{sec:hyperparameters}. This suggests that by effectively utilizing the limited knowledge of the environment (e.g., reverberation level and rough room geometry), the SANN model can yield more robust filters than the traditional methods. However, when measured data is available and only stationary PSZ rendering is required, the traditional methods may yield comparable or better performance than the SANN model as robustness is not a major concern in that case.

It is also worth comparing the actual implementation cost of the model to that of the traditional methods. The SANN model, once trained, requires only a forward pass of the neural network to generate the filters. Under the current experimental setup, the evaluated model has a total of 2.5 M parameters (which takes about 10 MB of memory) and takes about 0.65 ms on an Apple M1 Pro processor (using CPU only) for a single filter generation. In contrast, with the traditional methods, it would require either $\sim$5891 MB to store the pre-computed filters for all possible combinations of BZ and DZ positions (assuming a spatial grid with 5 cm resolution in the region of $[-1,1]$ m $\times [0.5,2]$ m), or significantly more computation resources to generate the filters on-the-fly as the ATF matrices are constantly changing with the positions of the zones. For example, a single filter generation based on the closed-form PM solution (Eq.~\ref{eq:PM_sol}) takes about 8.4 ms on the same CPU. The AM method would be even more computationally expensive as it requires iterative optimization for each filter generation.

\section{Summary and Final Remarks}\label{sec:summary}

In this work, we have proposed a deep learning-based approach that utilizes a spatially adaptive neural network (SANN) model for rendering head-tracked PSZs. Depending on whether measured ATFs are available, the SANN model can be trained either with simulated ATFs for robustness in unknown environments or with a mixture of simulated and measured ATFs for better isolation performance (i.e., IZI and IPI) in the measurement region. 

In evaluating the model's performance, we found that although the listeners can move freely within the rendering area, the isolation is fundamentally limited by the loudspeaker array size and the relative positions of the PSZs. We also found that the model is able to reduce the filter's artifacts (e.g., pre-ringing) by adding the time-domain loss term without significantly affecting the isolation performance, and that training with randomized room configurations yields better filter robustness compared to limiting the filter gains.

In comparison with traditional methods, we found that the SANN model yields similar or better isolation performance than the PM and AM methods with explicit regularization when no measured ATFs are available, and has fewer artifacts in the filter responses. The model is also more efficient in terms of computation and storage costs, making the real-time implementation of head-tracked PSZ rendering more feasible. 

Although the results in Sec.~\ref{sec:results} demonstrate the effectiveness of the proposed approach for head-tracked PSZ rendering, we emphasize that the model's architecture adopted in this work and its associated loss function are only an example of the possible design choices, and can be easily modified to meet other system requirements. For example, the model outputs can be set to be time-domain filter impulse responses, as done in \cite{pepe2022digital}, or to frequency-domain filter coefficients of different lengths; in addition, thanks to the flexibility of the loss calculation, other loss terms that are formulated in both the frequency and time domains can also be added to the model to further improve the performance. Moreover, the proposed approach applies also to a wide range of other audio rendering tasks that require head tracking or position-dependent processing, such as crosstalk cancellation, loudspeaker equalization, and sound field synthesis.

To simplify the analysis, we made an assumption that the actual listeners are absent in both filter generation and evaluation. Although the acoustic scattering effects due to the listener's head are not dominant in the frequency range of interest (100-1500 Hz), the model's performance may be significantly affected by torso-related effects and the inter-listener scattering effect \cite{qiao2023effects} at higher frequencies in realistic scenarios. Therefore, the model should further incorporate listener-specific information, such as the head-related transfer functions (HRTFs) and the anthropometric data of the listener, to achieve better isolation performance at higher frequencies. 

As suggested by the results, the bottleneck of the model's performance is the lack of measured ATFs for training. Therefore, future work should include developing more efficient data collection methods and spatial sampling schemes or improving the simulation method to better represent the listening environment when acoustic measurements are not feasible. Furthermore, the balance between the robustness inside and outside the measurement region should also be further optimized.

\section*{Acknowledgments}
This work was supported by a research grant from Masimo Corporation. The authors acknowledge the use of the Princeton Research Computing resources at Princeton University which is consortium of groups led by the Princeton Institute for Computational Science and Engineering (PICSciE) and Office of Information Technology's Research Computing.

\bibliographystyle{IEEEtran}
\bibliography{refs_ieee}

\newpage

 


\vfill

\end{document}